# Comment on "Effects of transverse magnetic field on mixed convection in wall plume of power-law fluids", by Rama Subba Reddy Gorla, Jin Kook Lee, Shoichiro Nakamura and Ioan Pop [International Journal of Engineering Science, 31 (1993) 1035-1045]

Asterios Pantokratoras

School of Engineering, Democritus University of Thrace,
67100 Xanthi – Greece
e-mail:apantokr@civil.duth.gr

In the above paper the authors treat the boundary layer mixed convection flow of a power-law fluid along a vertical adiabatic surface in a transverse magnetic field with a steady thermal source at the leading edge. The governing non-similar equations are solved by means of a novel finite difference scheme. However, there are two fundamental errors in this paper and the presented results do not have any practical value. This argument is explained below:

1. The momentum equation used by the authors is

$$u\frac{\partial u}{\partial x} + v\frac{\partial u}{\partial y} = \frac{K}{\rho}\frac{\partial}{\partial y}\left[\left|\frac{\partial u}{\partial y}\right|^{n-1}\frac{\partial u}{\partial y}\right] + \varepsilon g\beta(T - T_\infty) - \frac{\sigma B^2 u}{\rho} \qquad (1)$$

where u and v are the velocity components, K is the consistency index, ρ is the fluid density, n is the power-law index, ε is +1 for buoyancy assisted flow, -1 for buoyancy opposed flow, T is the fluid temperature, g is the gravity acceleration, β is the fluid thermal expansion coefficient, σ is the fluid electrical conductivity and B is the strength of the applied magnetic field. The boundary conditions are:

at y = 0: u = 0, v = 0, $\partial T/\partial y = 0$     (2)
as y → ∞   u = $U_\infty$, T = $T_\infty$     (3)

where $U_\infty$ is the free stream velocity. Let us apply the momentum equation at large y. At large distances from the plate the fluid temperature is equal to ambient temperature and the buoyancy term gβ(T-T$_\infty$) in momentum equation is zero. Taking into account that, at large distances from the plate, velocity is everywhere constant and equal to U$_\infty$ the velocity gradient ∂u/∂y is also zero. The same happens with the diffusion term (first term on the right hand side of the momentum equation). This means that the momentum equation takes the following form at large y

$$U_\infty \frac{\partial U_\infty}{\partial x} = -\frac{\sigma B^2}{\rho} U_\infty \qquad (4)$$

or

$$\frac{\partial U_\infty}{\partial x} = -\frac{\sigma B^2}{\rho} \qquad (5)$$

From the above equation we see that the free stream velocity should change along x and therefore the momentum equation (1) is not compatible with the assumption that the free stream velocity is constant. This is an error made frequently in the literature (see Vajravelu, 2007 and Xu, 2007).

2. The energy equation used in the above paper is

$$u \frac{\partial T}{\partial x} + v \frac{\partial T}{\partial y} = a \frac{\partial}{\partial y}\left[\left|\frac{\partial u}{\partial y}\right|^{n-1} \frac{\partial T}{\partial y}\right] \qquad (6)$$

where α is the fluid thermal diffusivity. However there is no scientific evidence that the energy equation of a power-law fluid should be different from that of a usual Newtonian fluid. There are numerous papers in the literature concerning heat transfer in non-Newtonian power-law fluids where the energy equation used is the classical one

$$u \frac{\partial T}{\partial x} + v \frac{\partial T}{\partial y} = a \frac{\partial^2 T}{\partial y^2} \qquad (7)$$

See for example the works of Huang and Chen (1990), Wang (1994), Rao et al. (1999) and Zheng (2002) and page 392 in Kleinstreuer (1997).

Taking into account the above arguments it is clear that both the momentum and energy equation used are wrong and therefore the results presented in the paper by Gorla et al. (1993) are inaccurate.

# REFERENCES

1. R. S. R.Gorla, J. K. Lee, S. Nakamura and I. Pop, Effects of transverse magnetic field on mixed convection in wall plume of power-law fluids, International Journal of Engineering Science, 31 (1993), 1035-1045.
2. K. Vajravelu, Corrigendum to " Convective heat transfer in an electrically conducting fluid at a stretching surface with uniform free stream", International Journal of Engineering Science 45(2007) 185-186.
3. H. Xu, Erratum to "An explicit analytic solution for convective heat transfer in an electrically conducting fluid at a stretching surface with uniform free stream", International Journal of Engineering Science 45(2007) 716-717.
4. M. J. Huang and C.K. Chen, Local similarity solutions of free convective heat transfer from a vertical plate to non-Newtonian power law fluids, International Journal of Heat and Mass Transfer, Vol. 33 (1990), 119-125.
5. T.Y. Wang, Mixed convection heat transfer from a vertical plate to non-Newtonian fluids, International Journal of Heat and Fluid Flow, Vol. 16 (1994), 56-61.
6. J.H. Rao, D.R. Jeng and K. J. DeWitt, Momentum and heat transfer in a power-law fluid with arbitrary injection/suction at a moving wall, International Journal of Heat and Mass Transfer, Vol. 42 (1999), 2837-2847.
7. L.C. Zheng and X.X. Zhang, Skin friction and heat transfer in power-law fluid laminar boundary layer along a moving surface, International Journal of Heat and Mass Transfer, Vol. 45 (2002), 2667-2672.
8. C. Kleinstreuer (1997), Engineering Fluid Mechanics, Cambridge University Press, Cambridge.